\documentclass{article}%
\usepackage{amsmath}
\usepackage{chicago}%
\usepackage{amsfonts}%
\usepackage{amssymb}%
\usepackage{graphicx}
\begin{document}

\title{CHAOTIC ACCRETION IN A NON-STATIONARY ELECTROMAGNETIC FIELD OF A SLOWLY
ROTATING COMPACT STAR }
\author{BABUR M. MIRZA\\Department of Mathematics\\Quaid-i-Azam University, Islamabad. 45320. Pakistan. }
\maketitle

\begin{abstract}
We investigate charge accretion in vicinity of a slowly rotating compact star
with a non-stationary electromagnetic field. Exact solutions to the general
relativistic Maxwell equations are obtained for a star formed of a highly
degenerate plasma with a gravitational field given by the linearized Kerr
metric. These solutions are used to formulate and then to study numerically
the equations of motion for a charged particle in star's vicinity using the
gravitoelectromagnetic force law. The analysis shows that close to the star
charge accretion does not always remain ordered. It is found that the magnetic
field plays the dominant role in the onset of chaos near the star's surface.

\end{abstract}


\section{Introduction}

Accretion is a fundamental process by which stars accumulate matter from their
surroundings, it therefore plays an important role in stellar formation and
evolution. For normal stars accretion is principally due to the gravitational
field of the star. However for compact objects (classed as neutron stars,
pulsars, and white dwarfs) accretion is not only due to gravitational but also
electromagnetic effects. In vicinity of these stars matter is mostly in the
form of a highly conducting electron plasma, whereas the star itself is formed
of a degenerate highly conducting quasi-neutral plasma. The compact state of
the plasma inside the star is brought about by the intense gravitational field
of the star. Exterior to the star gravitational and electromagnetic field
produce combined effects on the dynamics of plasma surrounding the star. Here
gravitational field has a two-fold effect on charge accretion. Firstly it
effects the charge dynamics through the interaction of particle mass with the
gravitational field of the star. Secondly the background spacetime also
influences the electromagnetic field of the star which in turn effects the
particle orbit via charge interaction. In general, under such conditions, the
electromagnetic field of a compact star is not stationary but gains in
intensity as the star evolves. Observationally the magnetic field intensity
for a newly formed compact star to the final stages of a compact star can vary
from $10^{4}G$ to $10^{12}G$ and hence varies with time$^{[1-3]}$.

In this paper we investigate the combined effects of gravitational and a
non-stationary electromagnetic field on a charge particle dynamics in vicinity
of a compact star. The star is considered to be formed of a degenerate highly
conducting plasma. The magnetohydrodynamics (MHD) approximation is applied for
the plasma forming the star. Using a generalized definition$^{[4,5]}$ of the
electromagnetic field tensor for an MHD plasma in a curved spacetime the
Maxwell field equations are explicitly formulated for a non-stationary axially
symmetric electromagnetic field. Here the linearized Kerr metric is used to
describe the gravitational field outside the star with ZAMO (zero angular
momentum observers) as the stationary observer$^{[6]}$. The generalized
Maxwell equation are then solved for the electric and magnetic field
components using the separation of variable technique. These solutions are
then used to obtain the equations of motion (geodesic equations) in linearized
Kerr spacetime for a charge particle in the non-stationary electromagnetic
field of the star. The numerical solutions to these equations show that at
large distances from the star particle trajectories are well defined and
ordered, however as the particle falls towards the star its orbit cannot be
precisely predicted hence exhibits chaos. The solution show that star's
magnetic field plays the dominant role in bringing the charge accretion to a
chaotic state.

The layout of the paper is as follows. In the next section generalized Maxwell
equations are formulated using the generalized electromagnetic field tensor.
These equations are solved in section 3 for a non-stationary axially symmetric
electromagnetic field using a separation of variable ansatz for the field
components. Then in section 4 equations of motion are formulated and
numerically solved for a charged particle lying in the field of the star. In
the conclusion (section 5) numerical results of section 4 are interpreted and
the main results of the study are summarized. Throughout we have used the
gravitational units in which $G=1=c$ unless mentioned otherwise.

\section{Formulation of the Electromagnetic Field Equations}

General relativistically the gravitational field of a rotating star can be
described by the Kerr metric. In the region exterior to a slowly rotating
compact star of mass $M$ a linearization of the Kerr metric is given by%

\begin{equation}
ds^{2}=-e^{2\Psi(r)}dt^{2}-2\omega(r)r^{2}\sin^{2}\theta dtd\varphi+r^{2}%
\sin^{2}\theta d\varphi^{2}+e^{-2\Psi(r)}dr^{2}+r^{2}d\theta^{2},\
\end{equation}
where we have used the usual Boyer-Lindquist coordinates. Also in expression
(1) $e^{2\Psi(r)}$ and $\omega(r)$ are given by the relations $e^{2\Psi
(r)}=(1-2M/r),\omega(r)\equiv d\varphi/dt=-g_{t\varphi}/g_{\varphi\varphi}$,
where $\omega(r)$ is the angular velocity of a free falling frame brought into
rotation by the frame dragging of the spacetime$^{[7]}$. The four velocity
components $u_{r}$ and $u_{\theta\text{ }}$ of a stationary observer ZAMO
circling the star at a fixed radial distance vanish, therefore using
$u^{\alpha}u_{\alpha}=-1$, the components of the four velocity vector for a
ZAMO are:%

\begin{equation}
u^{\alpha}=e^{-\Psi(r)}(1,0,0,\omega(r)),\quad u_{\alpha}=e^{\Psi
(r)}(-1,0,0,0).
\end{equation}
For a star formed of highly degenerate plasma the electromagnetic field
$(E^{\alpha},B^{\alpha})$ of the star can be determined by the Maxwell
equations. In a curved spacetime, such as the Kerr spacetime the general
relativistic form of the Maxwell equations is given by%

\begin{equation}
F_{\alpha\beta,\gamma}+F_{\beta\gamma,\alpha}+F_{\gamma\alpha,\beta}=0,
\end{equation}%
\begin{equation}
\left(  \sqrt{-g}F^{\alpha\beta}\right)  _{,\beta}=4\pi\sqrt{-g}J^{\alpha},
\end{equation}
where $g$ represents the determinant of the metric tensor $g_{\alpha\beta}$.
Here $F_{\alpha\beta}$ is the generalized electromagnetic field tensor for an
ideal magnetohydrodynamic fluid given by a unique tensor expression:%

\begin{equation}
F_{\alpha\beta}=u_{\alpha}E_{\beta}-u_{\beta}E_{\alpha}+\eta_{\alpha
\beta\gamma\delta}u^{\gamma}B^{\delta},
\end{equation}
and $J^{\alpha}$ is current four vector. In general the current four vector is
the sum of two terms corresponding to a convection and to a conduction
current: $J^{\alpha}=\epsilon u^{\alpha}+\sigma u_{\beta}F^{\beta\alpha} $,
where $\epsilon$ is the proper charge density and $\sigma$ is the conductivity
of the plasma. The volume element 4-form $\eta_{\alpha\beta\gamma\delta}$ is
equal to $\sqrt{-g}\epsilon_{\alpha\beta\gamma\delta}$ where $\epsilon
_{\alpha\beta\gamma\delta}$ is the Levi -Civita symbol. Since in a compact
star the quasi-neutral degenerate plasma is essentially a perfect conductor,
we assume that for the region external to the star $J^{\alpha}$ can be
approximated to zero (see references [8] and [9] for a discussion of this
often made assumption for a quasi-neutral degenerate plasma).

To determine the non-stationary electromagnetic field of the star we assume
that the electric and magnetic field components are dependent not only on the
polar coordinates $r$ and $\theta$ but also the time coordinate $t$ in the
frame of ZAMO. We then obtain from the Maxwell equations (3) and (4) the
following determining equations for the electromagnetic field in the slow
rotation approximation$^{[10]}$:%
\begin{equation}
\partial_{r}(r^{2}\sin\theta u^{t}B_{{}}^{r})+\partial_{\theta}(r^{2}u^{t}%
\sin\theta B^{\theta})=0,
\end{equation}%
\begin{equation}
r^{2}\sin\theta u^{t}\partial_{t}B^{\theta}-\partial_{r}(u_{t}E_{\varphi})=0,
\end{equation}%
\begin{equation}
r^{2}\sin\theta u^{t}\partial_{t}B^{r}+u_{t}\partial_{\theta}E_{\varphi}=0,
\end{equation}%
\begin{equation}
r^{2}\sin\theta u^{t}\partial_{t}B^{\varphi}+\partial_{r}(u_{t}E_{\theta
}-r^{2}\sin\theta u^{\varphi}B^{r})-\partial_{\theta}(u_{t}E_{r}-r^{2}%
\sin\theta u^{\varphi}B^{\theta})=0,
\end{equation}%
\begin{equation}
\partial_{r}(r^{2}\sin\theta u^{t}E^{r})+\partial_{\theta}(r^{2}u^{t}%
\sin\theta E^{\theta})=0,
\end{equation}%
\begin{equation}
r^{2}\sin\theta u^{t}\partial_{t}E^{\theta}{}-\partial_{r}(u_{t}B_{\varphi
})=0,
\end{equation}%
\begin{equation}
r^{2}\sin\theta u^{t}\partial_{t}E^{r}{}+u_{t}\partial_{\theta}B_{\varphi}=0,
\end{equation}%
\begin{equation}
r^{2}\sin\theta u_{{}}^{t}\partial_{t}E^{\varphi}+\partial_{r}(u_{t}B_{\theta
}-r^{2}\sin\theta u^{\varphi}E^{r})-\partial_{\theta}(u_{t}B_{r}-r^{2}%
\sin\theta u^{\varphi}E^{\theta})=0.
\end{equation}

\section{Solution for the Electromagnetic Field Exterior to the Star}

The non-stationary electromagnetic field of the star is now determined by the
system of equations (6) to (13). To obtain the time dependent electric and
magnetic field components $E^{\alpha}$ and $B^{\alpha}$ we assume the
following separation ansatz for the field components:%

\begin{equation}
E^{r}=\mathcal{A}(r)\mathfrak{a}(t)\frac{\alpha}{\sin^{2}\theta},\qquad
E^{\theta}=\mathcal{B}(r)\mathfrak{b}(t)\frac{\beta}{\sin\theta},\qquad E
_{\substack{^{{}} \\}}^{\varphi}=\mathcal{C}(r)\mathfrak{c}(t)\frac{\gamma
}{\sin^{2}\theta}%
\end{equation}%

\begin{equation}
B^{r}=\mathcal{D}(r)\mathfrak{d}(t)\frac{\delta}{\sin^{2}\theta},\qquad
B^{\theta}=\mathcal{E}(r)\mathfrak{e}(t)\frac{\varepsilon}{\sin\theta},\qquad
B_{\substack{^{{}} \\}}^{\varphi}=\mathcal{F}(r)\mathfrak{f}(t)\frac{\lambda
}{\sin^{2}\theta}%
\end{equation}
Substituting into (6) we obtain%

\begin{equation}
\frac{d}{dr}(r^{2}u^{t}\ \mathcal{D})=0
\end{equation}
This implies that $\mathcal{D}(r)=const./r^{2}u^{t}$. Also from equation (10)
we obtain $\mathcal{A}(r)=const./r^{2}u^{t}$ hence the radial components of
the electric and magnetic field reduce as:
\begin{equation}
E^{r}(r,\theta,t)=\delta\mathfrak{d}(t)/r^{2}u^{t}\sin^{2}\theta,\quad
B^{r}(r,\theta,t)=\alpha\mathfrak{a}(t)/r^{2}u^{t}\sin^{2}\theta
\end{equation}
From (13) it therefore follows that%

\begin{equation}
r^{2}u^{t}\mathcal{C\gamma}\frac{d\mathfrak{c}}{dt}+\frac{d}{dr}(r^{2}%
u_{t}\mathcal{E}\mathfrak{e}\varepsilon-r^{2}\omega u^{t}\mathcal{A}%
\mathfrak{a}\alpha)-u_{t}\sin\theta\frac{dB_{r}}{d\theta}=0
\end{equation}
Comparing coefficients of $\sin\theta$ on both sides the last equation implies
that\ $B_{r}$ must be independent of $\theta$. This is compatible with
equation (17) iff $\delta\equiv0$. Therefore it follows that $B^{r}=0$.
Similarly from (17)\ it follows that $E^{r}=0$. We therefore have after the
separation of variables, from equation (13), (11), (9), and (7) respectively,
the following determining equations%

\begin{equation}
\frac{1}{\mathfrak{e}}\frac{d\mathfrak{c}}{dt}=\frac{-\varepsilon}{\gamma
r^{2}u^{t}\mathcal{C}}\frac{d}{dr}(u_{t}r^{2}\mathcal{E})=k_{1}%
\end{equation}%

\begin{equation}
\frac{1}{\mathfrak{f}}\frac{d\mathfrak{b}}{dt}=\frac{\lambda}{\beta r^{2}%
u^{t}\mathcal{B}}\frac{d}{dr}(u_{t}r^{2}\mathcal{F})=k_{2}%
\end{equation}%

\begin{equation}
\frac{1}{\mathfrak{b}}\frac{d\mathfrak{f}}{dt}=\frac{-\beta}{\lambda
r^{2}u^{t}\mathcal{F}}\frac{d}{dr}(u_{t}r^{2}\mathcal{B})=-k_{3}%
\end{equation}%

\begin{equation}
\frac{1}{\mathfrak{c}}\frac{d\mathfrak{e}}{dt}=\frac{\gamma}{\varepsilon
r^{2}u^{t}\mathcal{E}}\frac{d}{dr}(u_{t}r^{2}\mathcal{C})=-k_{4}%
\end{equation}
where $k_{1}$, $k_{2}$, $k_{3}$, and $k_{4}$ are the separation constants.

\subsection{TIME DEPENDENT SOLUTIONS}

For the separated equations (19) to (22) we have for the time dependent parts
$\mathfrak{b(}t\mathfrak{)}$, $\mathfrak{c(}t\mathfrak{)}$, $\mathfrak{e(}%
t\mathfrak{)}$, and $\mathfrak{f(}t\mathfrak{)}$ the following set of
equations:%
\begin{equation}
\frac{d\mathfrak{c}}{dt}=k_{1}\mathfrak{e},\quad\frac{d\mathfrak{b}}{dt}%
=k_{2}\mathfrak{f}%
\end{equation}%
\begin{equation}
\frac{d\mathfrak{f}}{dt}=-k_{3}\mathfrak{b},\quad\frac{d\mathfrak{e}}%
{dt}=-k_{4}\mathfrak{c}%
\end{equation}
Differentiating the first of equations (23), and substituting in the second
equation of the pair (24) we obtain%

\begin{equation}
\mathfrak{c}(t)=A_{1}\exp(-i\sqrt{k_{1}k_{4}}t)
\end{equation}
Substituting back in equation (23) we obtain for $\mathfrak{e}$:%

\begin{equation}
\mathfrak{e}(t)=\frac{-i\sqrt{k_{1}k_{4}}}{k_{1}}A_{1}\exp(-i\sqrt{k_{1}k_{4}%
}t)
\end{equation}
where $A_{1}$ is a constant. In a similar manner we obtain for $\mathfrak{f}%
(t)$ and $\mathfrak{b}(t)$:%

\begin{equation}
\mathfrak{b}(t)=A_{2}\exp(-i\sqrt{k_{2}k_{3}}t)
\end{equation}%

\begin{equation}
\mathfrak{f}(t)=\frac{-i\sqrt{k_{2}k_{3}}}{k_{2}}A_{2}\exp(-i\sqrt{k_{2}k_{3}%
}t)
\end{equation}

\subsection{RADIAL SOLUTIONS}

The radial components $\mathcal{C}(r)$, $\mathcal{E}(r)$, $\mathcal{F}(r)$ and
$\mathcal{B}(r)$ are obtainable explicitly from equations (19) to (22) as
follows. First from the radial part of the separated equation (19) we have for
the component $\mathcal{C}(r)$:%

\begin{equation}
\mathcal{C}(r)=\frac{\varepsilon\sqrt{1-\frac{2M}{r}}}{k_{1}\gamma r^{2}%
}\frac{d}{dr}(r^{2}\sqrt{1-\frac{2M}{r}}\mathcal{E})
\end{equation}
Substituting into equation (22) gives an expression for $\mathcal{E}(r)$%

\begin{equation}
\frac{d}{dr}[(1-\frac{2M}{r})\frac{d}{dr}(\sqrt{1-\frac{2M}{r}}r^{2}%
\mathcal{E})]=\frac{k_{1}k_{4}r^{2}\mathcal{E}}{\sqrt{1-\frac{2M}{r}}}%
\end{equation}
This differential equation can be solved explicitly for the component
$\mathcal{E}(r)$ to give%

\begin{align}
\mathcal{E}(r)  & =r^{-3/2}(r-2M)^{-\frac{1}{2}+2M\sqrt{k_{1}k_{4}}}C_{1}%
\exp(\sqrt{k_{1}k_{4}}r)-\nonumber\\
& r^{-3/2}(r-2M)^{-\frac{1}{2}-2M\sqrt{k_{1}k_{4}}}\frac{C_{2}}{2\sqrt
{k_{1}k_{4}}}\exp(-\sqrt{k_{1}k_{4}}r)
\end{align}
The first part of the above solution is asymptotically unbounded, hence we
take only the second part of the solution for $\mathcal{E}(r)$. Putting then
for $\mathcal{E}(r)$ in equation (29) we obtain after some simplifications the
component $\mathcal{C}(r)$:%

\begin{equation}
\mathcal{C}(r)=\frac{-\varepsilon C_{2}\sqrt{1-\frac{2M}{r}}}{2\gamma
r^{9/2}\sqrt{k_{1}}k_{4}^{3/2}}(r-2M)^{-\frac{1}{2}-2M\sqrt{k_{1}k_{4}}%
}\{r(2+\sqrt{k_{1}k_{4}}r)-3M\}\exp(-\sqrt{k_{1}k_{4}}r)
\end{equation}
In a similar manner we obtain for $\mathcal{F}(r)$ and $\mathcal{B}(r)$:%

\begin{align}
\mathcal{F}(r)  & =r^{-3/2}(r-2M)^{-\frac{1}{2}+2M\sqrt{k_{2}k_{3}}}C_{3}%
\exp(\sqrt{k_{2}k_{3}}r)-\nonumber\\
& r^{-3/2}(r-2M)^{-\frac{1}{2}-2M\sqrt{k_{2}k_{3}}}\frac{C_{4}}{2\sqrt
{k_{2}k_{3}}}\exp(-\sqrt{k_{2}k_{3}}r)
\end{align}%
\begin{equation}
\mathcal{B}(r)=\frac{-\lambda C_{4}\sqrt{1-\frac{2M}{r}}}{2\beta r^{9/2}%
\sqrt{k_{2}}k_{3}^{3/2}}(r-2M)^{-\frac{1}{2}-2M\sqrt{k_{2}k_{3}}}%
\{r(2+\sqrt{k_{2}k_{3}}r)-3M\}\exp(-\sqrt{k_{2}k_{3}}r)
\end{equation}
These solutions are single valued and bounded for $r\rightarrow\infty$.

Hence the radial solutions (31) to (34) together with the time dependent
solutions (25) to (28) determine the electromagnetic field of a compact star
in the region exterior to the star.

\section{Charge Accretion in the Slow Rotation Approximation}

A linearization of the geodesic equation for a slowly rotating gravitational
source is given by the gravitoelectromagnetic force of magnitude
$\mathbf{G}+\mathbf{v}\times\mathbf{H}$ per unit mass where $\mathbf{G}$ and
$\mathbf{H} $ are the gravitoelectric (GE) and gravitomagnetic (GM)
potentials$^{[11,12]} $. In the slow rotation limit the star can be considered
as a homogeneous sphere of mass $M$ and radius $R$. In this case the GE and GM
potentials are given by%
\begin{equation}
\mathbf{G=}-\frac{M}{r^{2}}\hat{\mathbf{r}},\quad\mathbf{H}=-\frac{12}%
{5}MR^{2}(\mathbf{\Omega.r}\frac{\mathbf{r}}{r^{5}}-\frac{1}{3}%
\frac{\mathbf{\Omega}}{r^{3}}),
\end{equation}
where $\mathbf{\Omega}$ is the angular velocity vector of the star. In the
presence of an electromagnetic field the force law for a particle with mass
$m$ and charge $q$ is given by%

\begin{equation}
\frac{d^{2}\mathbf{r}}{dt^{2}}=(\mathbf{G}+\mathbf{v}\times\mathbf{H)+}%
\frac{q}{m}\mathbf{(E+v\times B).}%
\end{equation}
The equations of motion can be studied numerically in Cartesian coordinates by
transforming the polar coordinates into the Cartesian coordinates as
$x=r\sin\theta\cos\varphi,y=r\sin\theta\sin\varphi,z=r\cos\theta$. The
transformation gives for the electric field components $E^{x}=-r\sin\varphi
E^{\varphi},$ $E^{y}=r\cos\varphi E^{\varphi},$ $E^{z}=-r\sin\theta E^{\theta
},$ and similarly for the magnetic field $B^{x}=-r\sin\varphi B^{\varphi},$
$B^{y}=r\cos\varphi B^{\varphi},$ $B^{z}=-r\sin\theta B^{\theta}$. In a two
dimensional plane $\theta=\pi/2$ (the equatorial plane of the star), the
gravitomagnetic effects are maximum, here we get for the equations of motion
for the $(x,y)$ plane:%

\begin{equation}
\frac{d^{2}x}{dt^{2}}=-g-E^{x}+(H-B^{z})\frac{dy}{dt}=-g+yE^{\varphi
}+(H+yB^{\theta})\frac{dy}{dt}%
\end{equation}%
\begin{equation}
\frac{d^{2}y}{dt^{2}}=-g+E^{y}-(H-B^{z})\frac{dx}{dt}=-g+xE^{\varphi
}-(H+yB^{\theta})\frac{dx}{dt}\mathbf{,}%
\end{equation}
where from (14) and (15) we have for the real part of $E^{\varphi}$ and
$B^{\theta}$%

\begin{equation}
\operatorname{Re}E^{\varphi}=E_{0}\frac{\sqrt{1-\frac{2M}{r}}}{r^{9/2}%
}(r-2M)^{-\frac{1}{2}-2Mk}\{r(2+kr)-3M\}\exp(-kr)\cos kt
\end{equation}%

\begin{equation}
\operatorname{Re}B^{\theta}=B_{0}r^{-3/2}(r-2M)^{-\frac{1}{2}-2Mk}%
\exp(-kr)\sin kt
\end{equation}
where $r=\sqrt{x^{2}+y^{2}}$ , $k=\sqrt{k_{1}k_{4}}$, and $E_{0}%
\equiv-\varepsilon A_{1}C_{2}/2k_{1}k_{4}\sin^{2}\theta$ and $B_{0}%
\equiv\varepsilon A_{1}C_{2}/2k_{1}\sin\theta$ are constants for fixed value
of $\theta$, such as in the equatorial plane of the star. Also $g\equiv
\mid\mathbf{G}\mid=M/(x^{2}+y^{2}),$ and $H=\mid\mathbf{H}\mid=\mu
/(x^{2}+y^{2})^{3/2},$ whereas $\mu=(4/5R)\Omega$, $\Omega$ being the
magnitude of the angular velocity vector and $R$ is the radius of the star.
Transforming the electromagnetic field components in coordinates $(x,y)$ and
using expressions (37), (38) (39), and (40) we plot the trajectories of the
particle for various values of \ the parameters $M$, $\mu$, $E_{0}$, and
$B_{0}$ in figures (1) to (4); where we have taken $x(0)=2$, $y(0)=2$,
$dx/dt\mid_{t=0}=0$, $dy/dt\mid_{t=0}=0$ and $k=1.$

\section{Conclusions and Summary}

In this paper we have investigated charge accretion around a slowly rotating
compact star with a non-stationary electromagnetic field. In the region
exterior to the star the electromagnetic field is determined by solving the
generalized Maxwell equations (6) to (13). Then to study charge accretion in
vicinity of a compact star, the equations of motion for a charged particle in
the presence of the non-stationary electromagnetic field are solved
numerically. For the external region, where the solutions are stable and
convergent, our analysis shows that gravitational as well as electromagnetic
field determine the charge particle trajectories. However it is the magnetic
force that has the dominant effect on the transition from an ordered state of
motion to a chaotic one. Close to the star particle motion is particularly
sensitive to a change in the magnetic field of the star. Here an increase in
the gravitational field strength (gravitoelectric as well as gravitomagnetic)
directly causes a suppression in magnetically induced chaos in particle orbit
(Figure 3 and 4). The process however requires a substantial (about $10$ to
$100$ times) increase in the gravitoelectric field. In comparison a change in
the gravitomagnetic parameter $\mu$ from $2 $ to $10$ units introduces order
in particle's orbit. On the other hand the effects of electric field (Figure
2) on the charged particle motion are more complicated. As the electric field
becomes comparable to the magnetic field, the accretion trajectories become
precisely determined. Crossing this value, an increase in the electric field
causes the charged particle to reverse its direction of motion after which it
escapes from falling into the star. Magnetic field effects on charged particle
trajectory displayed in figure 1 also indicate that although an increase in
the magnetic field strength makes charge accretion chaotic, the general trend
is that of particle in-fall due to the dominant gravitational effects. On the
other hand in the regions close to the poles of the star, where magnetic field
is particularly high, it is expected that chaotic trend in accretion is more
dominant provided that the electric field effects on particle's motion are
comparatively weak.

Summarizing the above conclusions, accretion at relatively large distances
from the surface of a compact star is more or less ordered due to the dominant
effects of gravitational as well as electric fields. However in the star's
vicinity accretion becomes chaotic depending on the magnetic field strength
close to surface of the star as compared to the other forces involved, in
particular the gravitational and the electric forces.

\textbf{Acknowledgements}

I thank Dr. Hamid Saleem for useful comments. I also acknowledge Quaid-i-Azam
University Research Fund (URF-2006) for the financial support.

\textbf{Figure Captions;}

Figure 1: Effects of the magnetic field on charge particle trajectories. The
magnetic field strength varies from $B_{0}=10,100,1000,$ to $10000$ units,
while $M=1$, $\mu=1$, and $E_{0}=1$ in gravitational units.

Figure 2: Effects of the electric field on charge particle trajectories. The
electric field strength varies from $E_{0}=1,10,100,$ to $1000$ units, while
$M=1$, $\mu=1$, and $B_{0}=10$ in gravitational units.

Figure 3: Effects of the gravitoelectric field on charge particle
trajectories. The gravitoelectric field strength varies from $M=1,10,100,$ to
$1000$ units, while $\mu=1$, $E_{0}=1$ and $B_{0}=10$ in gravitational units.

Figure 4: Effects of the gravitomagnetic field on charge particle
trajectories. The gravitomagnetic field strength varies from $\mu=0.05,1,3,$
to $10$ units, while $M=1$, $E_{0}=1$ and $B_{0}=10$ in gravitational units.
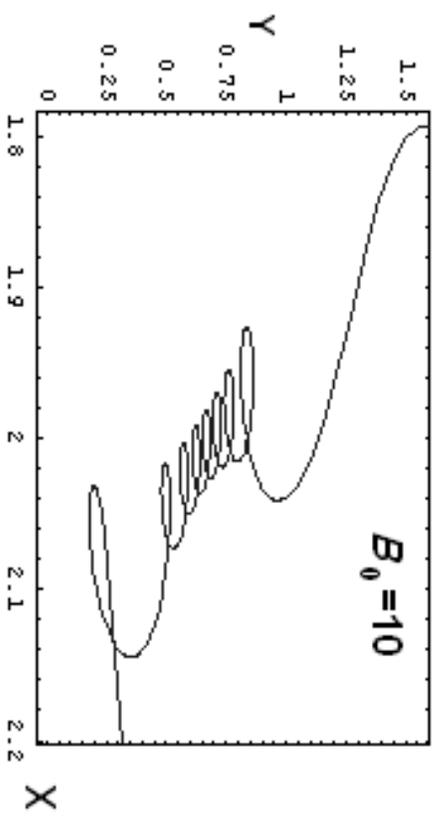
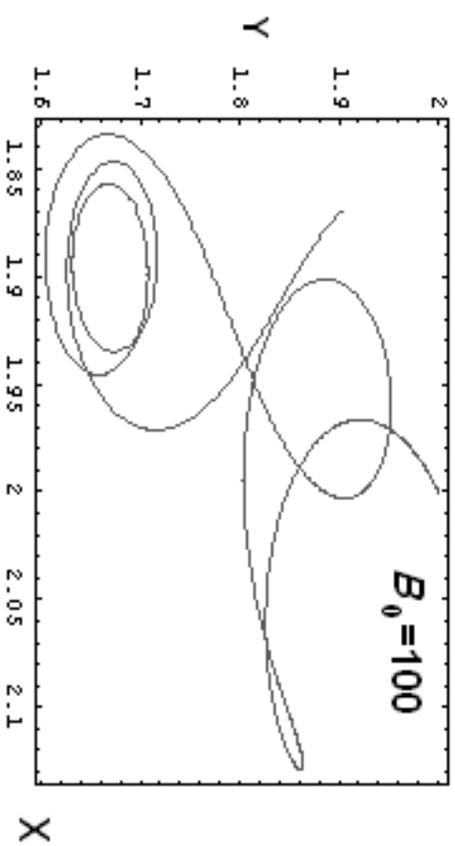
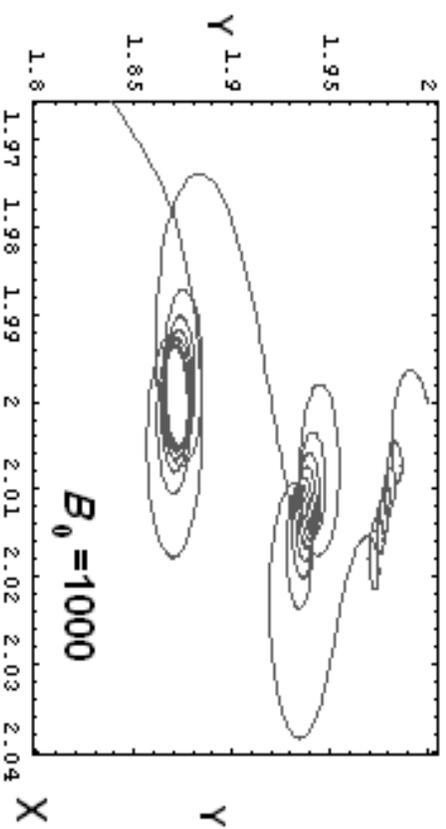
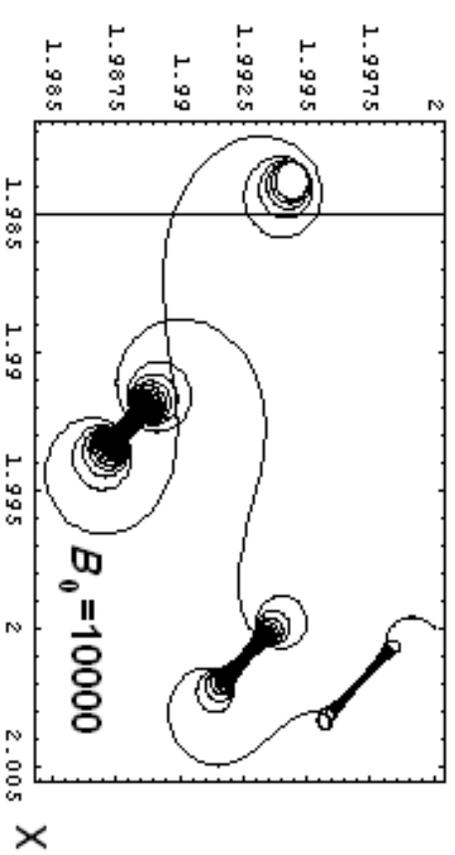
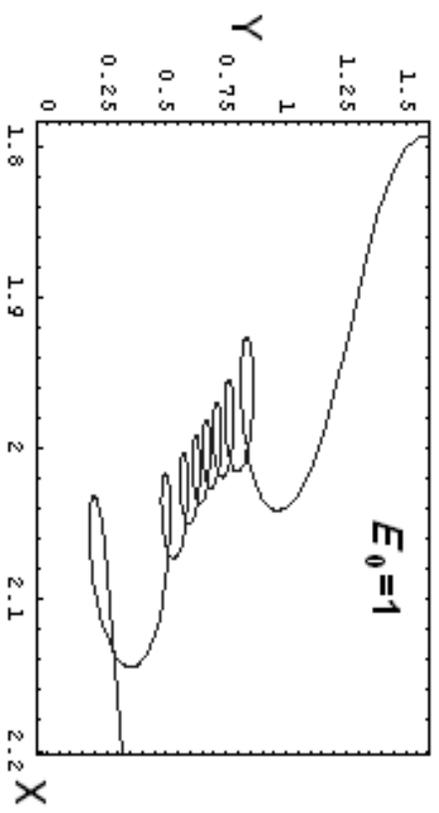
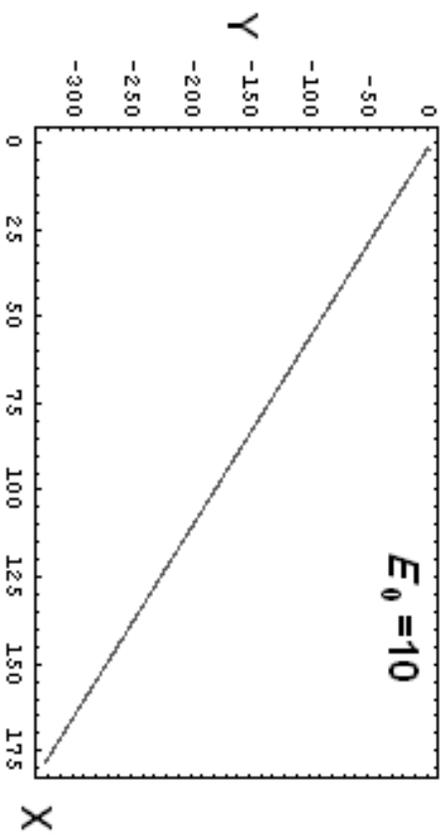
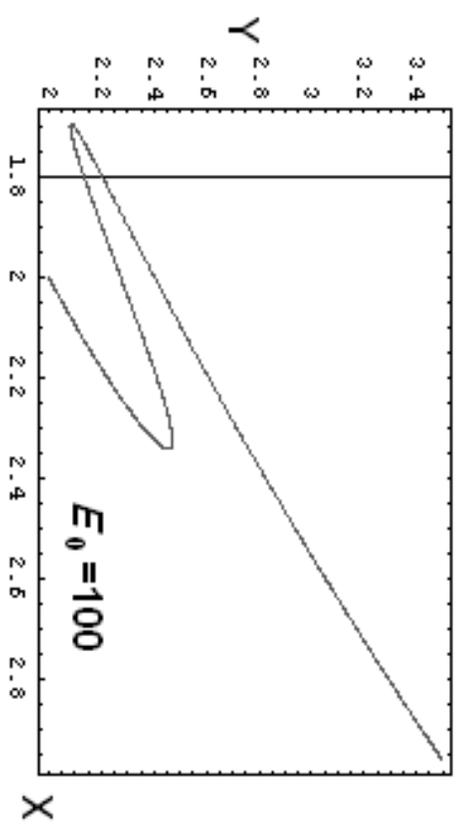
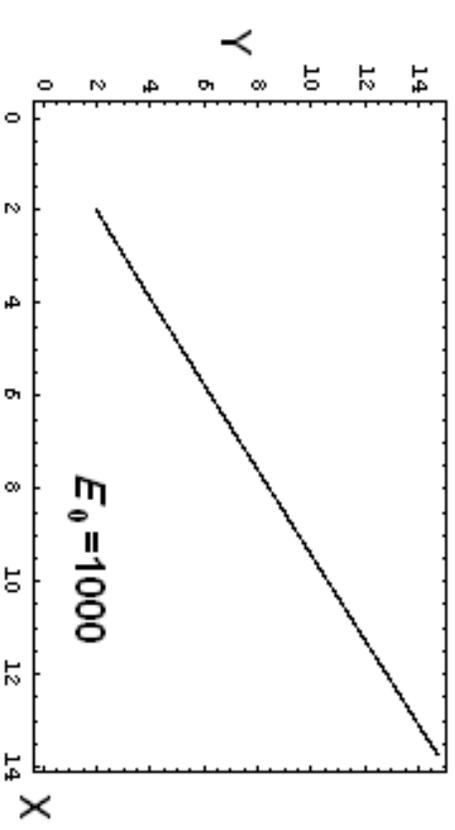
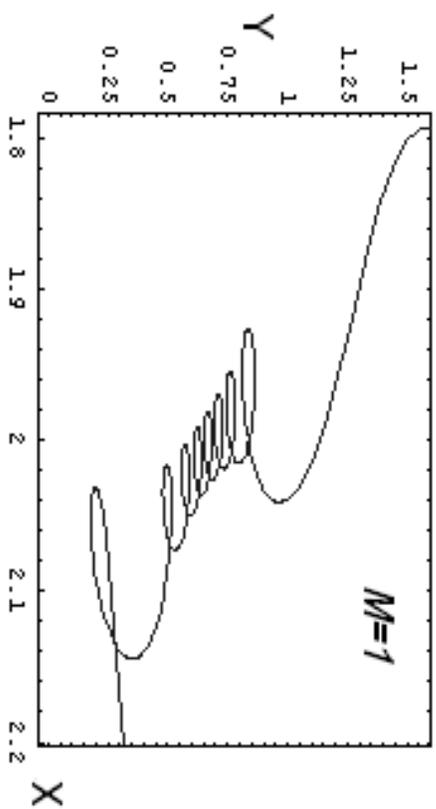
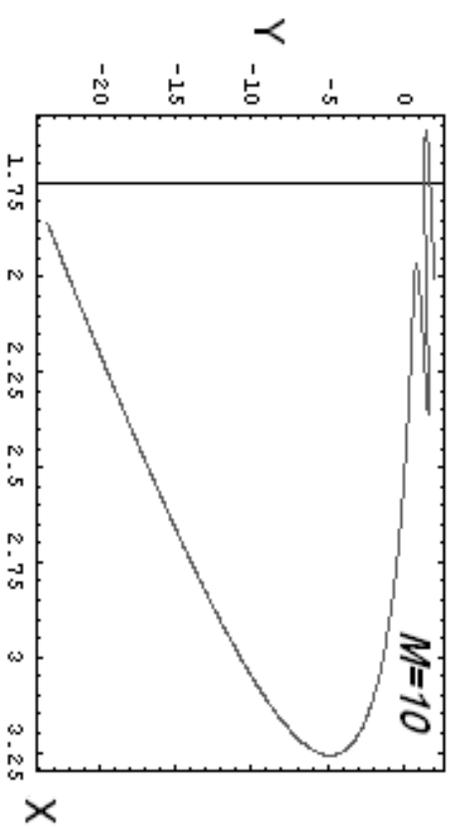
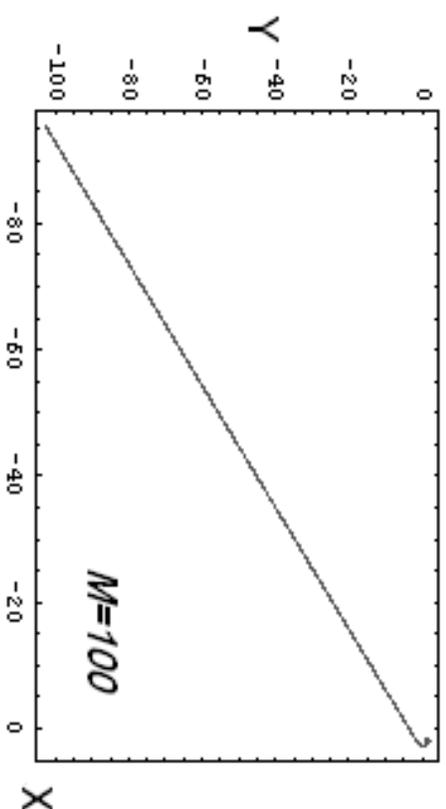
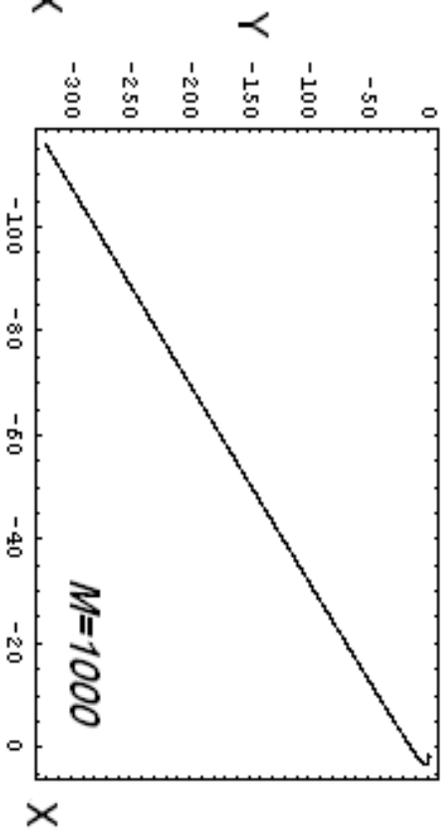
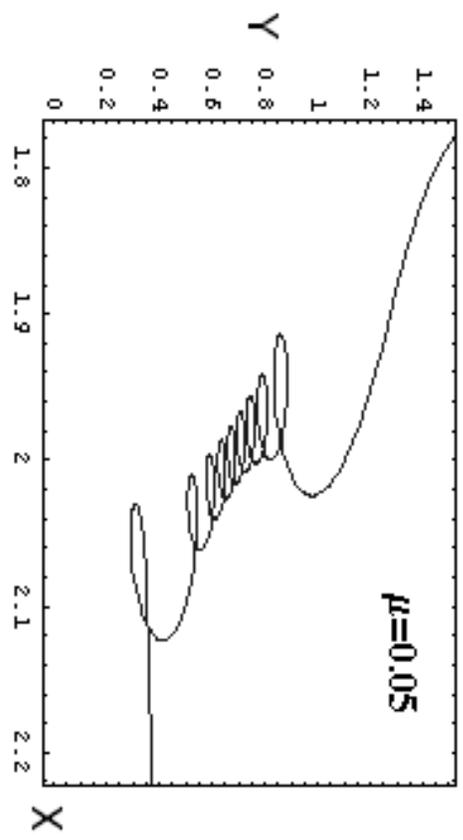
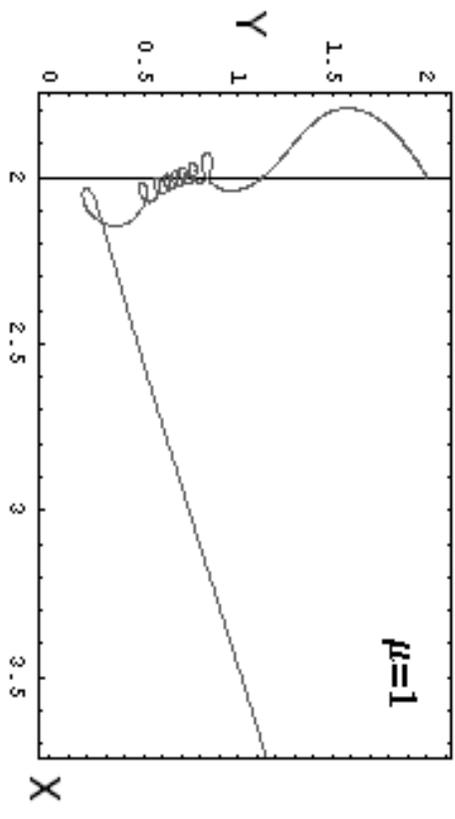
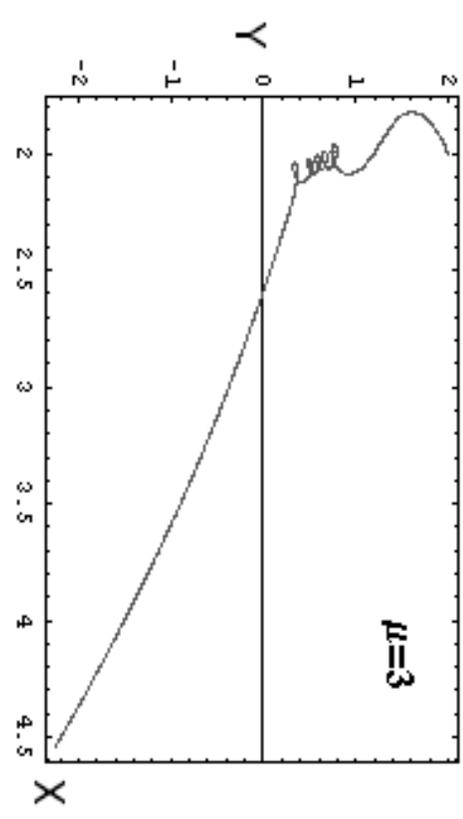
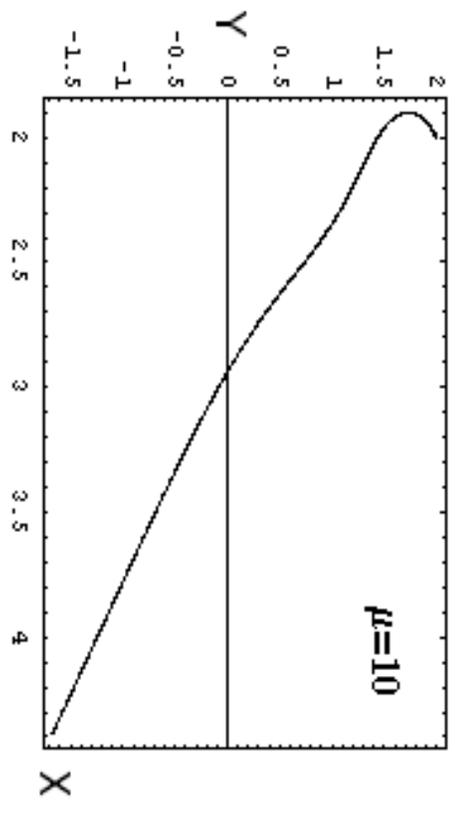
\end{document}